\documentclass[prl,aps,twocolumn]{revtex4}
\usepackage{amssymb}
\usepackage{amsmath}
\usepackage{graphicx}
\usepackage{t1enc}
\usepackage[utf8]{inputenc}

\begin{document}

\title{Probing the spacetime structure of vacuum entanglement}

\author{Andrzej Dragan}
\address{School of Mathematical Sciences, University of Nottingham, Nottingham NG7 2RD United Kingdom}
\address{Institute of Theoretical Physics, University of Warsaw, Ho\.{z}a 69, 00-049 Warsaw, Poland}
\author{Ivette Fuentes\footnote{Previously known as Fuentes-Guridi and Fuentes-Schuller.}}
\address{School of Mathematical Sciences, University of Nottingham, Nottingham NG7 2RD United Kingdom}

\begin{abstract}
We introduce a framework for probing the spacetime structure of vacuum entanglement that exhibits infinite range correlations between the future and the past, as well as between spatially separated regions. Our results are non-perturbative and strict.
\end{abstract}

\maketitle
Two independent quantum systems at rest localised in causally disconnected regions can extract entanglement from the vacuum state of a quantum field \cite{Reznik}. This striking result further investigated by several authors \cite{Massar,Lin:2008jj,Lin,Cliche} has evidenced that the vacuum state is entangled and perhaps one day may serve as an entanglement reservoir for relativistic quantum information protocols. Unfortunately, in such studies the entanglement rapidly decays as a function of the relative distance between the systems severely limiting the possibility of detecting the effect in the laboratory. Recently, it has been shown \cite{Olson:2011bq} that by independently quantising the past and future spacetime regions of a massless quantum field it is also possible to extract vacuum entanglement from timelike separated regions using quantum systems with an energy gap strongly varying in time - a situation difficult to realise in practice. The results mentioned above involved perturbative calculations, however the case of two spacelike separated counteraccelerating systems has been studied beyond the perturbative regime employing certain approximations \cite{Lin,Massar}. Again, the entanglement extracted decayed rapidly as the spacetime distance between the systems increased.

In this Letter we introduce a simple scheme to access both spacelike and timelike vacuum entanglement using a single extraction method which does not require the systems to have a time-dependent energy gap. Our scheme captures the essence of the previous studies providing further insights into the spacetime structure of vacuum entanglement: timelike and spacelike vacuum entanglement are indeed two aspects of the same phenomenon. We find that the entanglement extracted has a spatial and temporal periodic structure thus extending over the whole spacetime. In previous schemes the vacuum entanglement could only be extracted efficiently when the two systems are close to each other. Being able to easily avoid direct interactions between the systems by placing them very far apart gives us a clear experimental advantage over the other proposals. The infinite range of correlations we observe is a reflection of the known fact that the field propagator between two spatially separated points never vanishes regardless of how apart these points are. Our approach requires no approximations leading to exact results that go beyond the perturbative regime. 

Vacuum entanglement effects are commonly investigated using Unruh-DeWitt detectors which are point-like quantum systems locally interacting with a quantum field in flat \cite{Unruh, DeWitt, Crispino} or curved spacetime \cite{Louko:1998qf, Louko:2007mu}. These detectors which were introduced with the purpose of probing the Hawking-Unruh radiation have been recently found to have several interesting applications. For example, they can be used to distinguish between inertial and accelerated frames \cite{Dragan}, to distinguish between different spacetimes \cite{VerSteeg, Gibbons}, and to investigate the degradation of entanglement due to acceleration \cite{Lin:2008jj, Downes}. In the Unruh-DeWitt detector model the point-like system interacts with all the frequency field modes while in our approach the interaction is limited to a finite number of modes.  This scheme is a generalisation of the single-mode detector introduced in \cite{berryunruh} to measure the Unruh effect via a Berry phase. With this restriction we are able to apply well-known techniques used in quantum-optical continuous variables systems making possible an exact analysis without approximations nor perturbative methods. Different means can be used to assure that the detectors interact only with a finite number of modes. For example one can consider that each point-like system is contained within a small cavity transparent to a selection of modes, that effectively acts as a multi-frequency filter. 

Consider $n$ point-like detectors at rest labelled by the index $i$, each carrying an internal degree of freedom corresponding to a harmonic oscillator of  frequency $\omega_i$ described by the annihilation operator $\hat{d}_i$ or alternatively by position and momentum operators $\hat{q}_i$, $\hat{p}_i$, where: $\hat{d}_i = \frac{1}{\sqrt{2}}\left(\hat{q}_i+i\hat{p}_i\right)$. Each of the detectors couples locally to $m$ scalar field modes  of frequency $\Omega_j$ with corresponding annihilation operators $\hat{f}_j=\frac{1}{\sqrt{2}}\left(\hat{Q}_j+i\hat{P}_j\right)$. We consider $1+1$ dimensional spacetime and assume the minimum coupling between the detectors and the field given by the following Hamiltonian:
\begin{eqnarray}
\label{generalhamiltonian}
\hat{H}(t) &=& \sum_{i=1}^n\omega_i\left(\hat{d}_i^\dagger\hat{d}_i+\frac{1}{2}\right)+\sum_{j=1}^m\Omega_j\left(\hat{f}_j^\dagger\hat{f}_j+\frac{1}{2}\right)\nonumber \\ \nonumber \\
& &+\sum_{i,j}\lambda_{ij}(t)(\hat{d}_i+\hat{d}_i^\dagger)\left(\hat{f}_j e^{i \Omega_j x_i}+\hat{f}_j^\dagger e^{-i \Omega_j x_i}\right)
\nonumber \\
&=& \sum_{i=1}^n\frac{\omega_i}{2}\left(\hat{q}_i^2+\hat{p}_i^2\right)+\sum_{j=1}^m\frac{\Omega_j}{2}\left(\hat{Q}_j^2+\hat{P}_j^2\right)\\ \nonumber \\
& &+2\sum_{i,j}\lambda_{ij}(t)\hat{x}_i\left(\hat{X}_j \cos(\Omega_j x_i)-\hat{P_j}\sin(\Omega_j x_i)\right)\nonumber,
\end{eqnarray}
where $\lambda_{ij}(t)$ are coupling coefficients and $x_i$ is the classical spatial coordinate of the $i$-th detector. The time evolution operator $\hat{U}(t)$ for the above Hamiltonian can be computed analytically, without any approximations if the coupling coefficients $\lambda_{ij}$ are time-independent making the time-ordering of the evolution operator trivial: $\hat{U}(t)=e^{-iHt}$. Since the Hamiltonian \eqref{generalhamiltonian} is quadratic and can be written as $\hat{H} = \frac{1}{2}\hat{X}^T W \hat{X}$, where $\hat{X}^T = (\hat{x}_1, \hat{p}_1,\ldots, \hat{x}_n, \hat{p}_n,\hat{X}_1, \hat{P}_1,\ldots, \hat{X}_m, \hat{P}_m)$ and $W$ is a real and symmetric matrix, the evolution of any position or momentum operator is given by a linear combination of all the operators involved. One can prove the following formula \cite{symplectic}:
\begin{equation}
\label{theorem}
e^{-\frac{i}{2}\hat{X}^T W \hat{X}t}\hat{X}e^{\frac{i}{2}\hat{X}^T W \hat{X}t} = e^{iKWt}\hat{X},
\end{equation}
where $K_{ij}=[\hat{X}_i, \hat{X}_j]$. When the Hamiltonian \eqref{generalhamiltonian} is time independent we can apply Eq. \eqref{theorem} since the operators acting on the vector $\hat{X}$ on the left-hand-side of the equation can be identified with the evolution operator $\hat{U}(t)$ and its hermitian conjugate. Therefore, the evolution of any position or momentum operator is given by the action of the linear symplectic operator $S=e^{iKWt}$ on $\hat{X}$. If the initial state of the system is Gaussian then the above evolution operator transforms it into another Gaussian state that can be fully characterised by a covariance matrix $\sigma_{ij} = \langle \hat{X}_{i}\hat{X}_{j}+\hat{X}_{j}\hat{X}_{i}\rangle-2\langle\hat{X}_{i}\rangle\langle\hat{X}_{j}\rangle$ \cite{Gerardo}. Although this formalism holds in the general case, our analysis focuses on detectors coupled to a single-frequency mode. 

Let us apply this scheme to study the entanglement acquired by a pair of resting detectors placed at positions $\pm x$, coupled resonantly to the same single frequency field mode $\omega$. We assume that the oscillators are initially in their ground states and the field is in the vacuum state, which is an overall Gaussian state. The entanglement can be quantified by the negativity ${\cal N}(\sigma)$ which is a function of the symplectic invariants of the covariance matrix $\sigma$ \cite{Gerardo}. It has been proven that ${\cal N}$ is a measure of entanglement when the state is Gaussian and the necessary and sufficient condition for inseparability of the state characterised by $\sigma$ is ${\cal N}(\sigma)>0$ \cite{Simon}.

\begin{figure}
\includegraphics[width=\columnwidth]{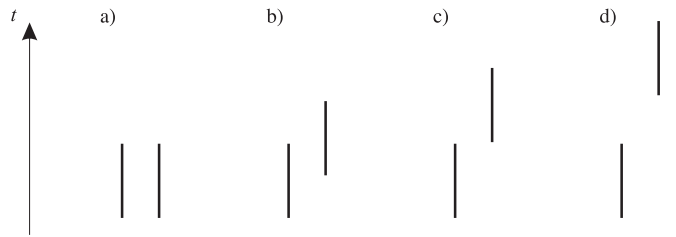}
\caption{\label{couplings}Detectors' configurations considered. The vertical lines represent trajectories of the detectors interacting with the field.}
\end{figure}

\begin{figure}
\includegraphics[width=\columnwidth]{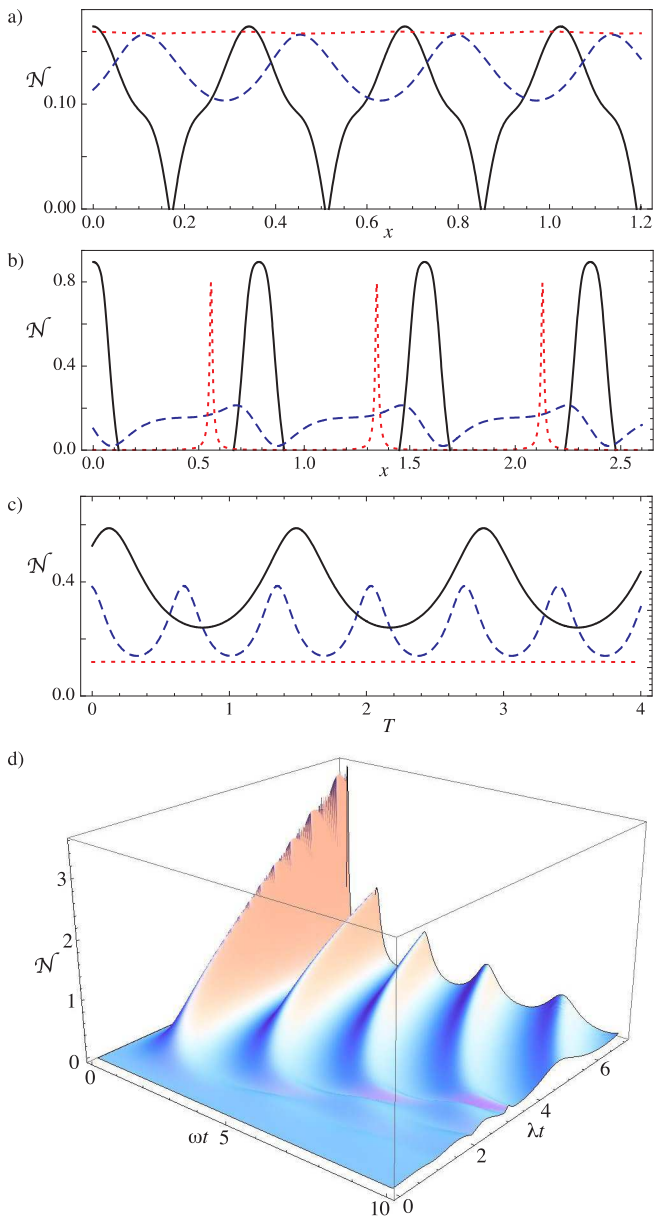}
\caption{\label{doublestraight}Entanglement ${\cal N}$ between the two resting detectors a) as a function of the separation $x$ for fixed $\omega=4.6$, $\lambda=1.5$, when both detectors are simultaneously turned on for a time $t=1$ (solid line); the second detector turned on for a time $t=1$ with a delay equal to $t=1/2$ with respect to the first detector (dashed line); the second detector turned on for a time $t=1$ immediately after the first one has been switched off (dotted line); c) as a function of delay $T$ introduced before the second detector starts interacting with the field (after the first one is off), for fixed $\omega=2.3$, $\lambda=1.2$, $t=1$, and $x=0$ (solid line), for fixed $\omega=4.6$, $\lambda=1.9$, $t=1$, and $x=0$ (dashed line), for fixed $\omega=4.6$, $\lambda =1.4$, $t=1$, and $x=0$ (dotted line); d) for $x=\frac{\pi}{2\omega}$ and two detectors simultaneously turned on for a time $t=1$ as a function of $\omega$ and $\lambda$.}
\end{figure}

We investigate the following scenarios: a) both detectors simultaneously turned on for a time $t$; b) the second detector turned on for a time $t$ with a delay equal to $t/2$ with respect to the first detector; c) the second detector turned on for a time $t$ immediately after the first one has been switched off; d) the same as c) but with an extra delay $T$ before turning on the second detector with $x=0$ - see Fig.~\ref{couplings}. All the interactions involve coupling coefficients given by time-dependent step functions. Defining $S_{0}(t)$ to be a $6\times 6$ symplectic matrix describing the free evolution of the system, $S_i(t)$ the evolution of the system when the $i$-th detector interacts with the field for a time $t$ while the other detector is switched off, and $S_{12}(t)$ when both detectors are simultaneously on, the overall evolution of the system in the four cases is given by: a) $S_{12}(t)$; b) $S_1(t/2)S_{12}(t/2)S_2(t/2)$; c) $S_1(t)S_2(t)$; d) $S_1(t)S_0(T)S_2(t)$. Using this technique we obtain results that are strict and hold for both spatially and temporally separated detectors. We find that for any choice of $\omega$, $\lambda$ and $t$ the entanglement changes periodically as the distance between the systems increases, however the behavior strongly depends on the specific values of the parameters. Examples of this are shown in Fig.~\ref{doublestraight}, where we plot the negativity ${\cal N}$ for the first three cases as a function of the detectors' separation $x$ for $\omega=4.6$, $\lambda=1.5$, $t=1$ (Fig.~\ref{doublestraight}a) and for $\omega=2$, $\lambda=4.8$, $t=1$ (Fig.~\ref{doublestraight}b).

It is evident that this setting allows one to extract both spacelike and timelike entanglement. It is interesting to compare our results to those obtained in \cite{Reznik} where the problem is investigated using the standard Unruh-DeWitt coupling in a particular fourth-order perturbation expansion valid only for two causally disconnected detectors. These authors show that entanglement rapidly vanishes with the increasing relative distance between the detectors. One may conjecture that by limiting the field frequency bandwidth one can increase the spacetime range of the correlations acquired by the detectors. The last case is plotted in Fig.~\ref{doublestraight}c) for fixed $\omega$, $\lambda$, $t$ and $x=0$ as a function of $T$. This situation corresponds to the entanglement acquired by the two detectors, one in the past and another in the future, as studied in \cite{Olson:2011bq}. Since in our case $S_0(T)$ is periodic in time, the resulting correlations will also be periodic. We find that timelike entanglement appears in the standard Minkowski quantisation and can be naturally extracted for massless or massive fields without the need of introducing a time-dependent frequency gap between the energy levels of the detectors as opposed to \cite{Olson:2011bq}. Our calculation shows that spacelike and timelike entanglement are two aspects of the same phenomenon. 

\begin{figure}
\includegraphics[width=\columnwidth]{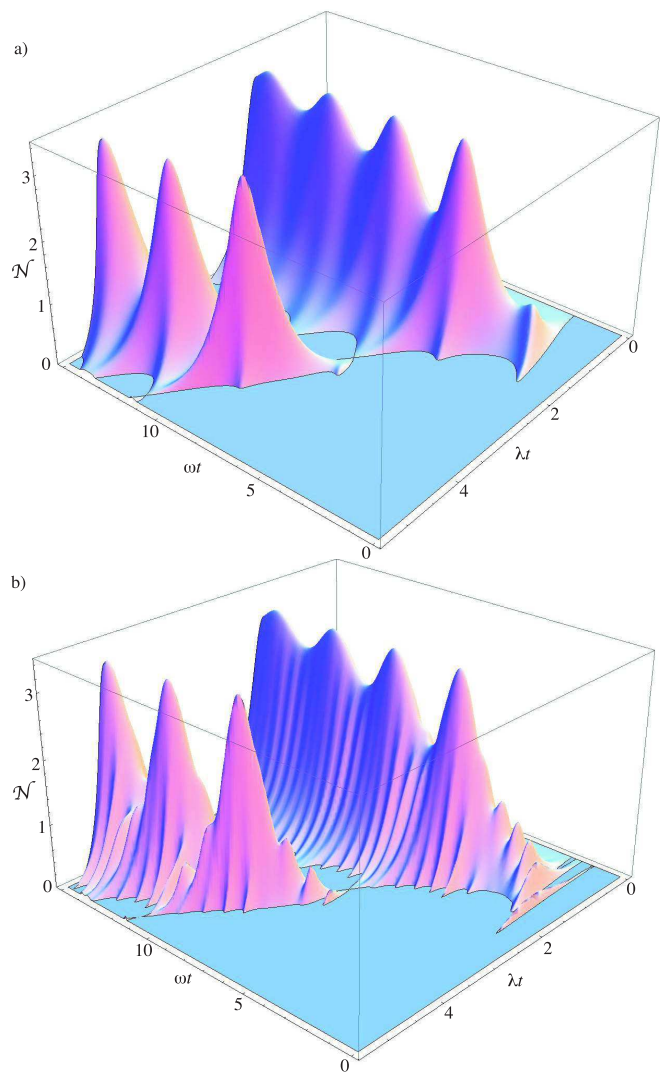}
\caption{\label{doubleaccelerated} Entanglement ${\cal N}$ between two detectors moving with the opposite accelerations corresponding to the squeezing parameter $r=1$ a) for the case where the detectors are simultaneously turned on at the Minkowski time $t=0$ and simultaneously turned off at the Minkowski time $t=1$; b) when an extra delay of $t=4$ has been introduced before the second detector starts interacting with the field.}
\end{figure}

Finally, we plot in Fig.~\ref{doublestraight}d) the entanglement ${\cal N}$ as a function of $\omega t$ and $\lambda t$, for fixed $x=\frac{\pi}{2\omega}$ when the detectors are simultaneously on. The result shows how to adjust the coupling coefficients and frequencies, as well as the time of the interaction, in order to maximise the entanglement acquired by the detectors. 

Our technique can be also applied to study the entanglement acquired by a pair of uniformly counteraccelerated detectors following two mirror trajectories corresponding to fixed positions in the Rindler coordinate system \cite{Massar,Lin}. The first/second detector moves in the Rindler wedge I/II with an acceleration $\pm a$ coupling to a single-frequency Rindler mode $\Omega$. This mode is analogous (via the Einsteinian Principle of Equivalence) to a gravitationally shifted plane wave near the horizon of a static black hole. The Hamiltonian written in Rindler coordinates has also the form \eqref{generalhamiltonian} since the mode solutions of the scalar field have the same form in Minkowski and Rindler coordinates due to the conformzl invsriance of the Klein-Gordon equation. The main difference in the calculation is that the detectors are coupled to Rindler frequency modes having support in separate Rindler regions. The initial state of the field in this case must be written in the Rindler basis in which it has a form of a two-mode squeezed state of the Rindler vacuum $|\text{vac}_M\rangle = \hat{\cal S}(r)|\text{vac}_R\rangle$ \cite{Unruh}. Here $|\text{vac}_M\rangle$ and $|\text{vac}_R\rangle$ are the vacuum states of the field in the Minkowski and Rindler frame of reference, respectively; $\hat{\cal S}(r)$ is the two-mode squeezing operator with $r$ varying with the acceleration as $\cosh(r)=(1-e^{-2\pi\Omega/a})^{-1/2}$. Since the squeezing operator $\hat{\cal S}(r)$ is also quadratic in the field mode operators, it can be written in the symplectic form using \eqref{theorem}. 

We plot our exact results in Fig.~\ref{doubleaccelerated}a) for the case where the detectors are simultaneously turned on at the Minkowski time $0$ and simultaneouly turned off at the Minkowski time $t$. We find that entanglement is observed in the weak and strong coupling regimes as long as $\lambda<\omega/2$. It is clear that this entanglement has been swapped (i.e. transfered) from the entanglement initially shared between the two Rindler modes. In Fig.~\ref{doubleaccelerated}b) an extra delay of $3t$ has been introduced before the second detector starts interacting. It is seen that contrary to the results of \cite{Massar,Lin}, the delay does not prevent us from extracting the same amount of entanglement as in the previous case. This is again due to the fact that in the case considered the free evolution operator is periodic in time.

Before making our concluding remarks we would like to include here a short analysis of a single detector interacting with a single field mode. The average number of detector excitations $N$ in the stationary case is given by:
\begin{eqnarray}
\label{inertialdetector}
N &=&  \frac{\lambda^2 t^2}{2}\left( \text{sinc}^2\left(\sqrt{\omega(\omega-2\lambda)}t\right)\right. \nonumber \\ \nonumber \\
& &+ \left.\text{sinc}^2\left(\sqrt{\omega(\omega+2\lambda)}t\right)\right).
\end{eqnarray}
In the uniformly accelerated case we find that the average number of excitations is modified in the following way:
\begin{eqnarray}
N(r) &=& N+R\sinh^2(r),
\end{eqnarray} 
where $N$ and $R$ depend only on the products $\omega t$ and $\lambda t$, and the former is given by the equation \eqref{inertialdetector}. One can see that the detector clearly reacts more vividly when accelerated, which is a clear signature of the Unruh effect.

In the emerging field of relativistic quantum information a first step to be taken is to find suitable ways to process information exploiting the relativistic aspects of quantum fields. In particular, the entanglement extracted from the vacuum by the detectors could be used as a resource for various information tasks. One might imagine using timelike entanglement to teleport information from a detector placed in the past to another one in a distant future. The scheme we introduce here allows for the implementation of such task. Its mathematical simplicity together with the possibility of manipulating point-like systems locally enables one to address a number of interesting questions and applications.

\section{Acknowledgements}
This work was presented at the Relativistic Quantum Information Workshop in the University of KwaZulu-Natal, South Africa, where we have received very useful feedback. We would like to thank the organisers, as well as G. Adesso, J. Doukas, N. Friis, A. Lee, J. Louko, S. Jay Olson, T. Ralph, A. Retzker for useful comments. I. F. thanks EPSRC  [CAF Grant EP/G00496X/2] for financial support.  

\bibliographystyle{apsrev}
\bibliography{references}

\end{document}